\pdfoutput=1

\documentclass[final,5p,twocolumn]{elsarticle}

\usepackage{amssymb}
\usepackage{hyperref}

\biboptions{sort&compress}

\begin{document}

\begin{frontmatter}

\title{The torque effect and  fluctuations  of  entropy deposition in rapidity\\ in ultra-relativistic nuclear collisions}

\author[agh]{Piotr Bo\.zek}
\ead{Piotr.Bozek@fis.agh.edu.pl}

\author[ifj,ujk]{Wojciech Broniowski}
\ead{Wojciech.Broniowski@ifj.edu.pl}

\address[agh]{AGH University of Science and Technology, Faculty of Physics and Applied Computer Science, 30-059 Krak\'ow, Poland}
\address[ifj]{The H. Niewodnicza\'nski Institute of Nuclear Physics, Polish Academy of Sciences, 31-342 Krak\'ow, Poland}
\address[ujk]{Institute of Physics, Jan Kochanowski University, 25-406 Kielce, Poland}

\date{8 June 2015}

\begin{abstract}

The decorrelation of the orientation  of the  event-plane  angles  
in the initial state of relativistic Pb-Pb and p-Pb
 collisions, the ``torque effect'',
 is studied in a model of entropy deposition in the longitudinal direction involving fluctuations 
of the longitudinal source profile on large scales.
The radiation from a single wounded nucleon is asymmetric in space-time rapidity. It is assumed that 
the extent in rapidity of the region of deposited entropy is random. 
Fluctuations in the deposition of  entropy from each source increase the event-plane decorrelation:
for Pb-Pb collisions they improve the description of the data, while for p-Pb collisions the mechanism is 
absolutely essential to
 generate any sizable decorrelation. We also show that the experimental data for rank-four flow may be 
 explained via folding of the elliptic flow. The results 
suggest the existence of 
long range fluctuations in the space-time
 distribution of entropy in the initial stages of relativistic nuclear collisions.
\end{abstract}


\end{frontmatter}

\section{Introduction}

During the collective expansion of the  fireball formed in 
relativistic heavy-ion collisions  azimuthal deformations of the density
are transformed into azimuthal asymmetry of particle emission spectra~\cite{Gale:2013da,Heinz:2013th}. In the presence of collective flow,  the particle spectra contain the harmonic components
\begin{eqnarray}
\hspace{-3mm}\frac{dN}{p_\perp dp_\perp d\eta \, d\phi} 
&\propto&  \dots  + v_2(p_\perp,\eta)\cos[2(\phi-\psi_2)]\\
&& + \, v_3(p_\perp,\eta)\cos[3(\phi-\psi_3)] + \dots \ .  \nonumber 
\end{eqnarray}
In each collision, the event-plane of the
 second or third order harmonic flow is oriented predominantly along the 
direction of elliptic or triangular deformations of the fireball.
It has been suggested that the angles $\psi_n$  of the  event-plane orientation
might vary as a function of pseudorapidity~\cite{Bozek:2010vz}
or transverse momentum~\cite{Gardim:2012im}. The effect leads to the factorization 
breaking for the two-particle cumulant flow coefficients,
\begin{equation}
 V_{n\Delta}( t_1, t_2  ) <  \sqrt{V_{n\Delta}( t_1, t_1  ) V_{n\Delta}( t_2, t_2  )} \ ,
\end{equation} 
where $t_i$ is the transverse momentum or pseudorapidity,
\begin{eqnarray}
V_{n\Delta}(t_1, t_2)=\langle \langle \cos[n(\phi_1-\phi_2)] \rangle \rangle,
\end{eqnarray}
and the average is taken over  events and over  all particle pairs with particles $i$  in a  bin around $t_i$.

The factorization breaking in transverse momentum 
 has been studied quantitatively in dynamical 
models~\cite{Gardim:2012im,Heinz:2013bua,Kozlov:2014fqa} 
in p-Pb and Pb-Pb collisions. The hydrodynamic response from  fluctuating
initial conditions can describe the experimentally observed event-plane
 fluctuations and the factorization breaking in $p_\perp$~\cite{Aad:2014lta,Khachatryan:2015oea}.

\begin{figure}[tb]
\begin{center}
\vspace{-15mm}
\includegraphics[angle=0,width=0.35 \textwidth]{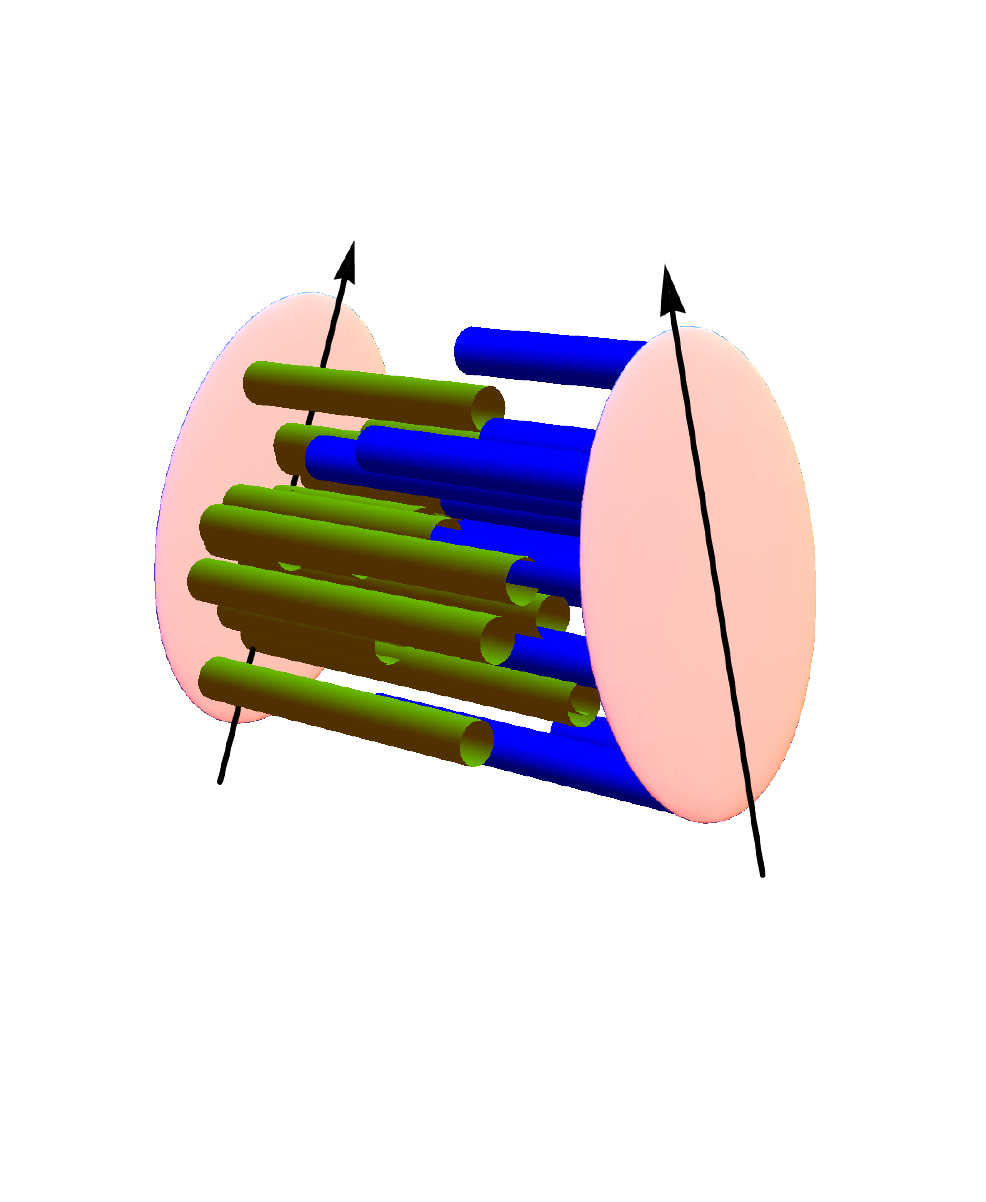}  
\end{center}
\vspace{-27mm}
\caption{ Schematic view of the entropy distribution in an early stage of an ultrarelativistic nuclear collision. The matter deposited from each
wounded nucleon occupies an interval in space-time rapidity with a randomly distributed end. As a result, the event-plane angles in the forward and 
backward bins are decorrelated.
\label{fig:pict}} 
\end{figure} 

The decorrelation of the event-plane angles at different pseudorapidities 
is seen in a number of calculations, both in hydrodynamic, cascade, or hybrid models~%
\cite{Bozek:2010vz,Petersen:2011fp,Xiao:2012uw,Jia:2014ysa,Jia:2014vja,Pang:2014pxa,Bozek:2015bha}. 
Nevertheless, a simultaneous description of the \mbox{Pb-Pb} and p-Pb data~\cite{Khachatryan:2015oea} poses 
a real challenge. In this paper we propose a decorrelation mechanism which is capable to grab the basic experimental features of both reactions.
A schematic view of the model is depicted in Fig.~\ref{fig:pict}, showing an early stage of the collision just after the two nuclei have passed through each other. 
The key ingredient is that the entropy deposition from the wounded nucleons~\cite{Bialas:1976ed} is made in string-like objects whose end-point is randomly distributed; some are 
longer and some shorter, with the length generated uniformly in the available rapidity interval. The idea is closely related to the model of Ref.~\cite{Brodsky:1977de}.

\section{The correlation measure}

It is very difficult to disentangle the genuine event-plane decorrelation due to the collective expansion of a  ``torqued'' fireball 
from non-flow fluctuations of short range in pseudorapidity~\cite{Bozek:2010vz}.
This difficulty is cleverly solved by using a factorization ratio using three bins
with a large separation in pseudorapidity, as proposed by the CMS collaboration~\cite{Khachatryan:2015oea}:
\begin{equation}
 r_n(\eta^a, \eta^b)=\frac{V_{n\Delta}(-\eta^a,\eta^b)}{V_{n\Delta}(\eta^a,\eta^b)}  \ , 
\label{eq:rn}
\end{equation}
with the forward reference bin $4,4< \eta_b < 5$ well  separated from the two 
central bins where $|\eta_a|<2.5$. The departure of the factorization ratio $r_n$ from unity is a measure of 
the event-plane angle decorrelation as a function of the pseudorapidity separation $\Delta \eta = 2 \eta_a$. 

In Ref.~\cite{Bozek:2015bha}, the factorization ratio for elliptic and triangular 
flow in Pb-Pb collisions at $\sqrt{s_{NN}}=2.76$~TeV is calculated in event-by-event viscous hydrodynamic simulations
 with Glauber initial conditions. Assuming an asymmetric entropy deposition in space-time  rapidity from left- and
right-going wounded nucleons, one finds that the orientation of the fireball deformation  depends on space-time
rapidity, as the contribution to the fireball entropy from target and projectile wounded nucleons 
changes with $\eta$~\cite{Bozek:2010vz}.  Calculations show that the event-plane decorrelation in pseudo-rapidity can
 be described qualitatively, but the factorization ratio is noticeably underestimated.
Moreover, the 
calculation cannot reproduce the observed factorization breaking in p-Pb collisions. 
 In the following, we discuss a mechanism introducing additional fluctuations in the entropy deposition, with long range correlations, that improves the description of the measured factorization ratio $r_n$.

In the presence of  collective expansion, the orientations of the event-planes and 
 the elliptic or triangular deformation are transformed into the orientation 
and the magnitude of the corresponding harmonic flow components~\cite{Gardim:2011xv}. By the same
mechanism, the torque of the event plane as a function of space-time rapidity is transformed into the
rapidity dependence of the event-plane orientation extracted from particle spectra. This relation is expected to hold for
decorrelation effects on large scales, while fluctuation in rapidity on small scales can be modified and 
washed out by the hydrodynamic evolution, resonance decays, mini-jets, etc. In the following,
we investigate a model of fluctuations in the entropy deposition in
space-time rapidity in the initial state. Hydrodynamic simulations show that the initial 
torque of the fireball in space-time rapidity is transformed into  a very similar torque in the 
pseudorapidity dependence of the harmonic flow event-planes~\cite{Bozek:2010vz,Bozek:2015bha}. 
Unfortunately, precise hydrodynamic calculations including non-flow effects are very demanding. 
In this paper, event-plane decorrelation in pseudorapidity for the second and third harmonic flow 
are approximated by the event-plane decorrelation in spacetime rapidity
 in the initial state.

Statistical hadronization, where a finite number of hadrons in a given bin is produced from a fireball with principal axes $\psi_n$, leads 
to large decorrelation effects~\cite{Bozek:2010vz} whose origin is trivial and needs to be canceled out. The CMS ratios (\ref{eq:rn}) accomplish this goal. 
Indeed, suppose we compute cumulants for the produced hadrons between the largely separated bins around $\eta_a$ and $\eta_b$. Then 
\begin{eqnarray}
&& \hspace{-14mm} V_{n\Delta}(\eta^a,\eta^b) = \langle \langle e^{i n(\phi_1-\phi_2)} \rangle \rangle = \langle \langle e^{i n(\psi_n(\eta_a)+\phi'_1-\psi_n(\eta_b)-\phi'_2)} \rangle \rangle \nonumber  \\
&& \simeq  \langle e^{i n[\psi_n(\eta_a)-\psi_n(\eta_b)]} \rangle \langle \langle e^{i n \phi'_1 -i n \phi'_2} \rangle \rangle, \label{eq:fact}
\end{eqnarray}
where the azimuths of the produced hadrons, $\phi_1$ and $\phi_2$, are evaluated in some reference frame, $\psi_n(\eta_a)$ and $\psi_n(\eta_b)$ are the event-plane angles of the fireball, and 
$\phi'_1$ and $\phi'_2$ are evaluated relative to $\psi_{n}(\eta_a)$ and $\psi_{n}(\eta_b)$, respectively. 
The factorization in Eq.~(\ref{eq:fact}) applies if the torque angle magnitude
 is uncorrelated with the flow magnitude.
The factors $ \langle \langle e^{i n \phi'_1-i n \phi'_2} \rangle \rangle$
cancel out in appropriate ratios. For the symmetric A-A collisions the production 
around $\eta_a$ is the same as around $-\eta_a$, hence taking the ratio (\ref{eq:rn}) accomplishes the goal. For asymmetric collisions, as p-A, 
the appropriate measure proposed by CMS is $\sqrt{r_n(\eta_a,\eta_b) r_n(-\eta_a,-\eta_b)}$.

According to the above discussion, the factorization ratio can be written as
\begin{equation}
r_n(\eta_a,\eta_b)=\frac{\langle \cos[ n(\psi_n(-\eta_a)-\psi_n(\eta_b))]\rangle}{\langle \cos[n(\psi_n(\eta_a)
-\psi_n(\eta_b))]\rangle} \ ,
\end{equation}
where the average is taken over events. Expanding $\psi_n(\pm \eta_a)\simeq \psi_n(0) \pm \frac{d\psi_n(\eta)}{d \eta}\eta_a$ yields
\begin{eqnarray}
& & \hspace{-14mm} r_n(\eta_a,\eta_b) \simeq  \\
& & \hspace{-14mm} \frac{\langle \cos[ n(\psi_n(0)\!-\!\psi_n(\eta_b))]\!-\!n \sin[ n(\psi_n(0)\!-\!\psi_n(\eta_b))] \frac{d\psi_n(0)}{d \eta}\eta_a\rangle}
              {\langle \cos[ n(\psi_n(0)\!-\!\psi_n(\eta_b))]\!+\!n \sin[ n(\psi_n(0)\!-\!\psi_n(\eta_b))]\frac{d\psi_n(0)}{d \eta}\eta_a\rangle}. \nonumber
\end{eqnarray}
For small values of the decorrelation angle, further expansion leads to
\begin{equation}
\hspace{-5mm} r_n(\eta_a,\eta_b) \simeq 1 - 2 n^2  \langle  (\psi_n(0)-\psi_n(\eta_b)))\frac{d\psi_n(0)}{d \eta}\rangle \eta_a \ .
\label{eq:rexp}
\end{equation}
The deviation of the factorization ratio from $1$  is found to be  approximately linear in $\eta_a$,
as observed by the CMS collaboration~\cite{Khachatryan:2015oea}. The deviation of the factorization ratio from $1$ 
in the initial state is given by the correlation of the twist angle and its derivative
\begin{eqnarray}
1-r_n & \simeq &   2n^2 \langle  (\psi_n(0)-\psi_n(\eta_b)))\frac{d\psi_n(0)}{d \eta}\rangle \eta_a \nonumber  \\
& \propto & \langle  (\psi_n(0)-\psi_n(\eta_b))^2\rangle \eta_a \ .
\label{eq:excor}
\end{eqnarray}
The last proportionality holds approximately because of the strong correlation between $\psi_n(0)-\psi_n(\eta_b)$
and $\frac{d\psi_n(0)}{d \eta}$. The slope $f_n$ of the linear dependence of 
\begin{equation} 
r_n(\eta_a,\eta_b)= 1 - 2 f_n \eta_a 
\label{eq:lin} 
\end{equation}
can be related to the variance of the event-plane angle difference between the central and the forward bin.
Due to event-by-event fluctuations, $\langle  (\psi_n(0)-\psi_n(\eta_b)))^2\rangle$
is found to be nonzero in several model calculations
 of the
 initial state~\cite{Bozek:2010vz,Petersen:2011fp,Xiao:2012uw}.
 
The $F^\eta_n$ parameter used by the CMS collaboration,
\begin{equation}
r_n(\eta_a,\eta_b) = e^{-2 F^\eta_n \eta_a},
\label{eq:F}
\end{equation}
 is approximately equal to the  slope $f_n$ of the linear dependence
(\ref{eq:lin}) for small factorization breaking. Parametrically $F^\eta_n \propto n^2$, which amplifies the
factorization breaking for higher harmonics. For centralities where the elliptic flow is 
strong one expects that the nonlinear contribution for $v_2^2$ dominates $v_4$. In 
that case $\psi_4 \simeq \psi_2$,
which leads to the relation 
\begin{eqnarray}
F^\eta_4/4\simeq F^\eta_2,
\label{eq:F4}
\end{eqnarray}
well satisfied in the experiment~\cite{Khachatryan:2015oea}.
Relation~(\ref{eq:excor}) cannot be easily applied to compare the size of $r_2$ and $r_3$, as the correlation between 
$\psi_n(0)-\psi_n(\eta_b)$ and $\frac{d\psi_n(0)}{d \eta}$ is stronger for $n=2$ than for $n=3$.

\section{Torque model with strings of fluctuating length}

The entropy distribution  in space-time rapidity in the initial stage is not yet fully understood.
In \mbox{3+1}-dimensional  hydrodynamic calculations, the  initial profile in the longitudinal direction is often 
assumed as a smooth symmetric function. This assumption is sufficient to obtain an average description
of pseudorapidity spectra in symmetric collisions. However, the radiation from forward- and backward-going color charges
naturally leads to asymmetric distributions in rapidity~\cite{Brodsky:1977de}. 
Following this idea, we assume a simple model where gluons radiated from a charge moving with rapidity $y_b$ 
are distributed in rapidity uniformly in a range $[y_a,y_b]$, with the end position $y_a$ taken as random (cf. Fig. \ref{fig:pict}).
When the distribution of  $y_a$ is uniform in the available range $[-y_{\rm beam}, y_{\rm beam}]$, which
is what we assume, then
the averaged distribution has a linear dependence on rapidity. Notably, such an approximately linear 
dependence of  the density of particles emitted from a single wounded nucleon  has been identified from particle spectra
in asymmetric d-Au collisions~\cite{Bialas:2004su}. The asymmetric linear (averaged) distribution is used 
successfully in the modeling of relativistic nuclear collisions~\cite{Adil:2005bb,Bozek:2010bi,Bzdak:2009dr}. 
Thus the model adopted by us to describe the fluctuations reproduces, upon averaging, the earlier approaches 
for observables computed from single rapidity bins.

Fluctuations in the distribution of right- and left-going nucleons give a torque in the event-plane orientation 
even in the average model~\cite{Bozek:2010vz}, that can partially reproduce  the factorization ratios 
$r_n(\eta_a,\eta_b)$ measured by the CMS collaboration. This average torque model predicts, as we shall see, a very small
factorization breaking in p-Pb collisions, unlike observed experimentally.
Let us  note that similar effects are expected in string models~\cite{Andersson:1986gw}, if 
rapidities of the color charges at the ends of the flux tube fluctuate. The investigated
mechanism  is restricted to fluctuations  which are long-range. 
The presence of any additional torque $\delta \psi_n(\eta)$ 
of the event-plane angles $\psi_n(\eta)+\delta \psi_n(\eta)$, coming from local clusters, thermalized 
jet remnants, etc., would not modify the factorization ratio $r_n(\eta_a,\eta_b)$ (see the discussion of the preceding Section), 
if the production in the forward and central bins is uncorrelated, $\langle \delta \psi_n(\pm \eta_a) \delta \psi_n(\eta_b) \rangle=0$.
The same argument applies for short-range non-flow correlations, as pointed out by the CMS collaboration~\cite{Khachatryan:2015oea}.

\begin{figure}
\begin{center}
\vspace{-5mm}
\includegraphics[angle=0,width=0.33 \textwidth]{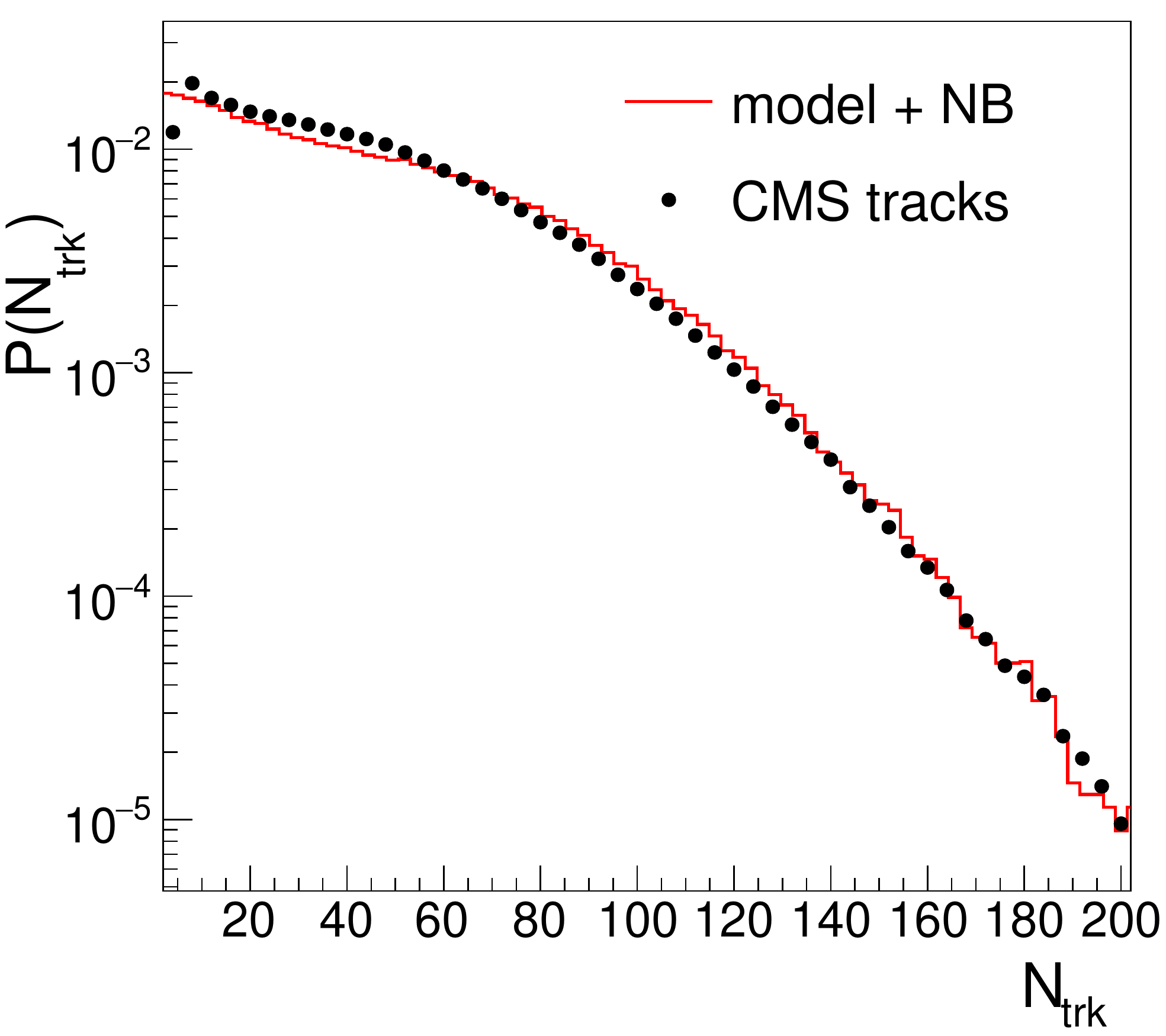}  
\end{center}
\vspace{-7mm}
\caption{Multiplicity distribution in p-Pb collisions, where the data 
are for charged tracks  with $p_\perp>0.4$~GeV and $|\eta|<2.4$ measured by CMS~\cite{cmswiki}, and the line 
denotes the corresponding results of the torque model with rapidity fluctuations, convoluted with a negative binomial distribution for the strength 
of the Glauber sources.
\label{fig:mult}} 
\end{figure}

The emergence of the torque effect relies on two features:
\begin{enumerate}
 \item asymmetric source profile in pseudorapidity, and 
 \item fluctuations.
\end{enumerate}
The fluctuations included in our model are two-fold. First, we incorporate the discussed fluctuations of the emission profile, as depicted in Fig.~\ref{fig:pict}, second, we 
fluctuate the strength of the sources, overlaying a suitable distribution over the distribution of the wounded nucleons. The combined amount of fluctuations is controlled by 
the multiplicity distributions. In particular, in p-Pb collisions we set the parameters of the overlaid distribution in such a way that we reproduce the CMS data in Fig.~\ref{fig:mult}.

Our calculations are carried out with {\tt GLISSANDO}~\cite{Broniowski:2007nz,Rybczynski:2013yba}. 
The realistic NN inelastic collision profile for the LHC energies is taken from Ref.~\cite{Rybczynski:2013mla}.
We use  an excluded distance $d=0.9$~fm when generating the nucleon configurations
in the nuclei. The total inelastic NN cross section is 64~mb for Pb-Pb collisions at $\sqrt{s_{NN}}=2.76$~TeV  and 70~mb  for p-Pb collisions at $\sqrt{s_{NN}}=5.02$~TeV.
The fireball density in the transverse plane and pseudorapidity $\eta$ is taken in a form as
a sum over the $N^+$ right-moving and $N^-$ left-moving wounded nucleons,
\begin{equation}
\hspace{-4mm} s(x,y,\eta)=\sum_{i=1}^{N^+} g^+_i(x,y,\eta) + \sum_{i=1}^{N^-} g^-_i(x,y,\eta)  .
\end{equation} 
The source density in the transverse plane involves a superposition of strength; $w_i$ is the superposed random weight, described in more detail below.  
We take into account the admixture of the binary collisions~\cite{Kharzeev:2000ph,Back:2001xy}. If $N^{\rm coll}_i$ denotes the number of collisions of the $i$-th nucleon with the nucleons from the other nucleus,
then $W^{\rm coll}_i=\sum_{j=1}^{N^{\rm coll}_i} w_j$ is the acquired random weight for the binary component. 
A necessary smearing is achieved with a smoothed Gaussian form centered around the position of the nucleon, $(x_i,y_i)$. 
Combining these elements yields
\begin{eqnarray}
g^\pm_i(x,y,\eta) &=& \left[w_i (1-\alpha)h^\pm(\eta) + W^{\rm coll}_i \alpha \right] H(\eta) \nonumber \\
 &\times& e^{ - \frac{(x-x_i)^2+(y-y_i)^2}{2\sigma^2}}. \label{eq:ggg}
\end{eqnarray}
The width of the smearing Gaussian is $\sigma=0.4$~fm, 
and the mixing parameter controlling the contribution of the binary collisions is $\alpha =0.15$.

The longitudinal density profile, according to the earlier discussion, has the form 
\begin{equation}
h_{i,\pm}(\eta)= 2 \theta[\pm(\eta_i -  \eta)], 
\label{eq:lprof}
\end{equation}
where $\theta$ denotes the step function and $\eta_i$ is distributed randomly in the range $[-y_{\rm beam}, y_{\rm beam}]$. 
Upon averaging over $\eta_i$, the function (\ref{eq:lprof}) yields ``triangular'' distributions 
used successfully in previous works.
The overall rapidity profile $H(\eta_\parallel)$ is taken 
as a plateau with Gaussian tails~\cite{Hirano:2002ds},
\begin{equation}
H(\eta)=\exp\left(-\frac{(|\eta|-\eta_p)^2\Theta(|\eta|-\eta_p)}{2\sigma_\eta^2}\right) ,
\end{equation}
where $\sigma_\eta=1.4$ and $\eta_p=2.5$~\cite{Bozek:2013uha}.

To summarize the above construction, the wounded nucleons lead to fluctuating and asymmetric production in pseudorapidity, 
while the binary collisions yield symmetric emission.

To set the parameters of the overlaid distribution producing the weights $w_i$ we proceed as in Ref.~\cite{Bozek:2013uha}. The production of charged
particles from each source of Eq.~(\ref{eq:ggg}) is described by the 
negative binomial distribution
\begin{equation}
N_{\lambda,\kappa}(n)=\frac{\Gamma(n+\kappa)\lambda^n\kappa^\kappa}{\Gamma(\kappa)n! (\lambda+\kappa)^{n+\kappa}}\ , \label{eq:NB}
\end{equation}
where the hadron multiplicity $n$ has the mean and variance given by  $\lambda$ and $\lambda(1+\lambda/\kappa)$, respectively. 
In Fig.~\ref{fig:mult} we show the result of the model fit to the CMS data~\cite{cmswiki}, where a very reasonable agreement in the large multiplicity tail is 
obtained. The optimum parameters are 
$\lambda=4.6$ and $\kappa=1.4$. We note that without the fluctuations in rapidity $\kappa=0.9$~\cite{Bozek:2013uha}, i.e., the variance of the fluctuations
of the strength of the sources must be larger in this case to reproduce the same distribution of hadrons. Assuming that the statistical hadronization following the deterministic hydrodynamic phase 
brings in an additional Poisson distribution for the number of hadrons, the
weights $w_i$ of 
entropy   of the Glauber sources follow the  $\Gamma$ distribution~\cite{Broniowski:2007nz}, 
\begin{equation}
P_\Gamma(w)=\frac{w^{\kappa-1}\kappa^\kappa}{\Gamma(\kappa)}e^{-\kappa w}\ 
 . \label{eq:Gamma}
\end{equation}

While the description of the model presented in this section seems rather involved, we note that apart for the fluctuations of the longitudinal extent of the sources from the wounded nucleons, 
which is novel and which upon averaging yields
the previously used emission profiles, the other elements (admixing binary collisions, overlaying the $\Gamma$ or negative binomial distributions) are standard in state-of-the art modeling of the 
Glauber phase of the collision, and the model parameters are fixed in the same way as in previous studies.

\begin{figure}[tb]
\begin{center}
\vspace{-15mm}
\includegraphics[angle=0,width=0.47 \textwidth]{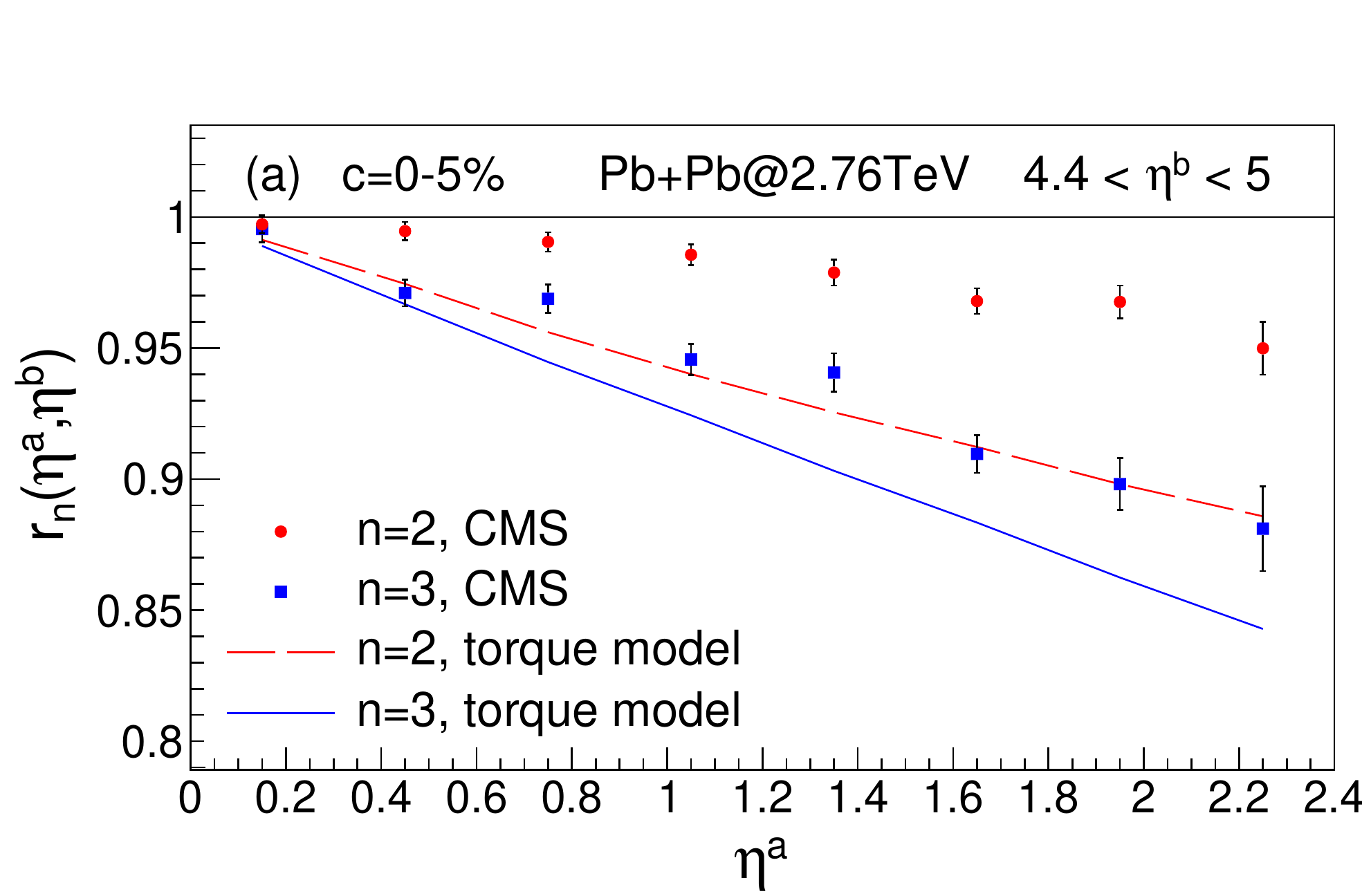}  \vspace{-12mm} \\
\includegraphics[angle=0,width=0.47 \textwidth]{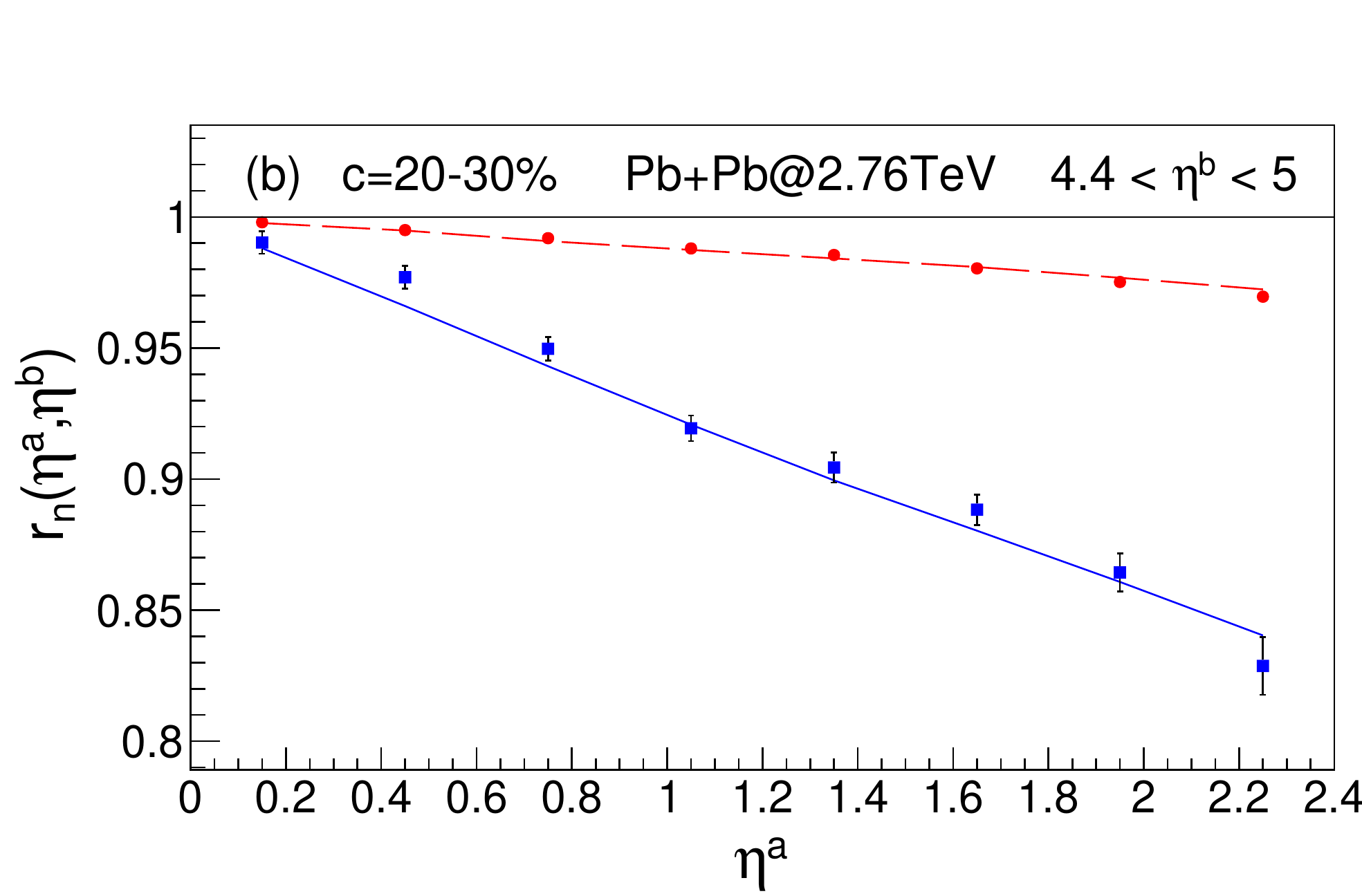}   \vspace{-12mm} \\
\includegraphics[angle=0,width=0.47 \textwidth]{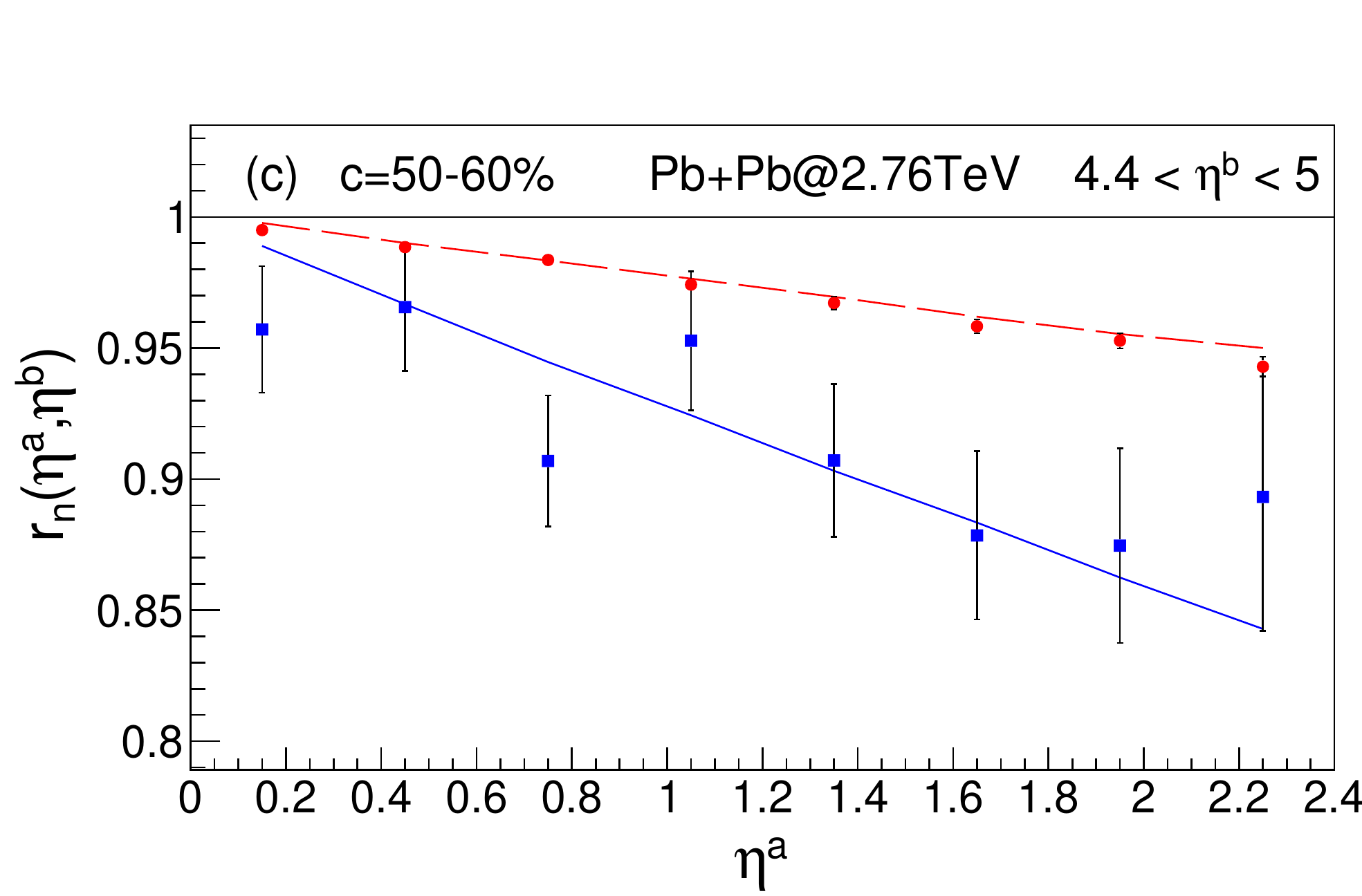}  
\end{center}
\vspace{-5mm}
\caption{Factorization ratios $r_n$ for the  elliptic and triangular flow, plotted
as functions of the central pseudorapidity bin position, obtained  from 
the torque model with fluctuating entropy distribution in rapidity (lines), and  the data of the CMS collaboration (symbols)~\cite{Khachatryan:2015oea}. 
In panels (a), (b) and (c) are presented results for centralities $0$-$5$\%, $20$-$30$\% 
and $50$-$60$\%, respectively.
\label{fig:rn}} 
\end{figure} 

\section{Results}

The factorization ratios for the second and third harmonic in Pb-Pb collisions are shown in Fig.~\ref{fig:rn}.
The calculation in the torque model with long-range fluctuations in pseudorapidity describes very well the data for
semi-central and peripheral collisions, both for $r_2$ and $r_3$.  We note, however, the lack of agreement
for $r_2$ in central collisions, where the decorrelation is significantly overestimated in the  model.
The result is nontrivial, as we were not able to adjust the emission profile used in the model to improve $r_2$ without spoiling the agreement for $r_3$.  
The data of the CMS collaboration show that the factorization ratio in central collisions depends on the choice of the reference pseudorapidity bin $\eta_b$~\cite{Khachatryan:2015oea}. 
This indicates that in central collisions fluctuations in the rapidity distribution of  short range or non-flow correlations 
become relatively more important. Such correlations could originate from hard physics that is outside 
our model of the initial stage. This issue calls for further studies.

\begin{figure}
\begin{center}
\vspace{-15mm}
\includegraphics[angle=0,width=0.47 \textwidth]{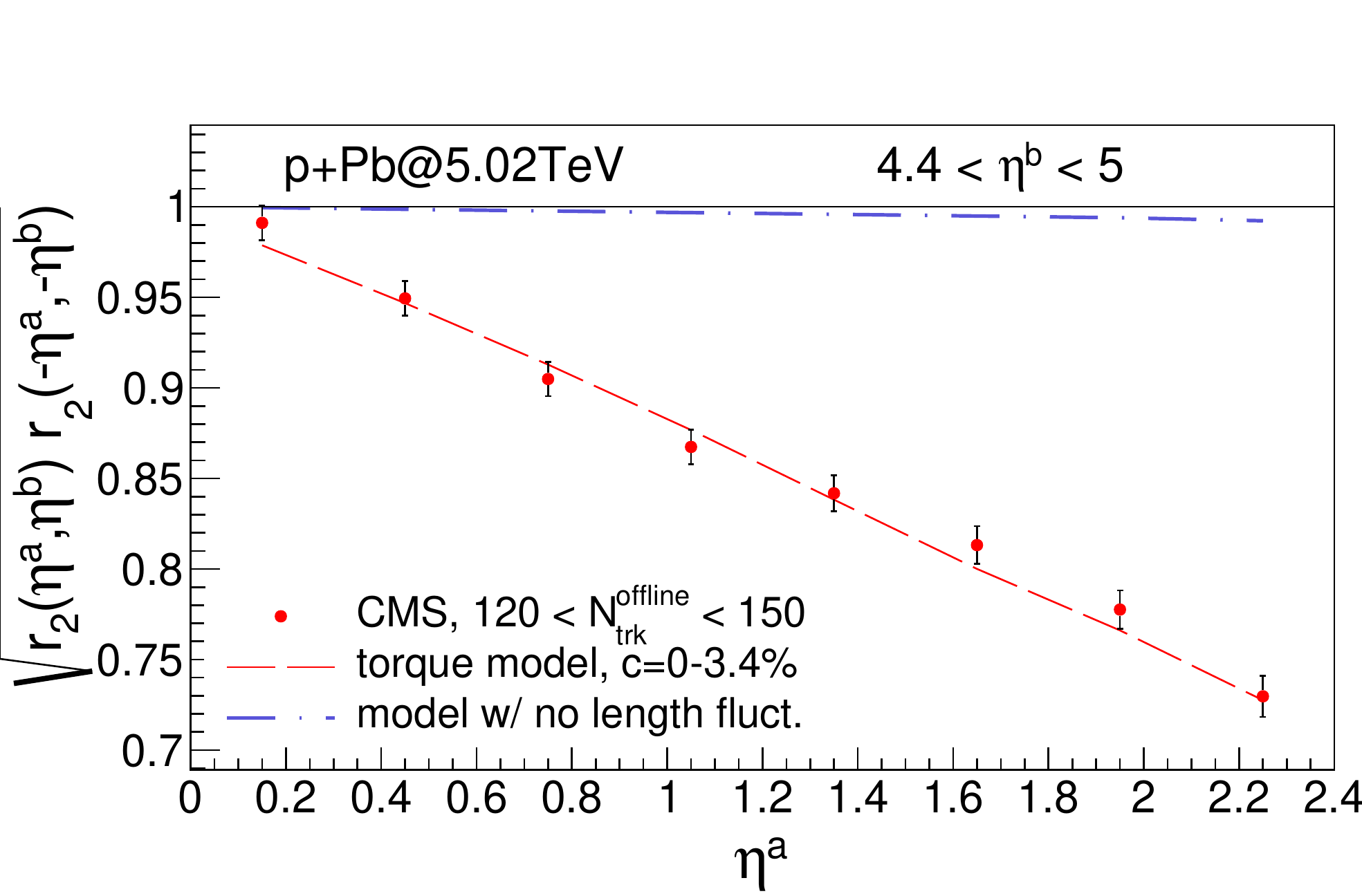}  
\end{center}
\vspace{-5mm}
\caption{Factorization ratio for the elliptic flow in p-Pb collisions, plotted
as a function of the central pseudorapidity bin position.
The data from the CMS collaboration (symbols)~\cite{Khachatryan:2015oea} agree well with the torque model with fluctuating entropy density in 
pseudorapidity (dashed line), while the model with averaged entropy profile in rapidity (dashed-dotted line) fails spectacularly. 
\label{fig:rpPb}} 
\end{figure}

The effect of fluctuations in the entropy distribution in rapidity is most striking for p-Pb collisions, as demonstrated in  
Fig.~(\ref{fig:rpPb}).  The experimental data show a significant factorization breaking in the second harmonic.
The calculation in the model with average entropy profile in rapidity (dashed-dotted line in Fig.~\ref{fig:rpPb})
gives almost no factorization breaking. The reason is simple, as the transverse profile at a given space-time
rapidity is dominated by the contribution from wounded nucleons from the Pb nucleus. When all these sources 
deposit the entropy in space-time rapidity in a similar way, the event-plane orientation will show almost no
rapidity dependence. The picture changes dramatically if the entropy density in space-time rapidity
for each source fluctuates, hence becomes different for each of them. Since the contribution of each  source from the Pb 
nucleus  to the entropy density at a given space-time rapidity varies, in consequence the event-plane orientation between the forward and central bin decorrelates noticeably. The calculation 
in the torque model with  fluctuating entropy distribution  in rapidity describes surprisingly well the experimental data (dashed line in Fig.~\ref{fig:rpPb}).  
This is the key result of our paper, which shows that the incorporation of the long-range pseudorapidity fluctuations is crucial to explain the large decorrelation seen in p-Pb collisions.

\begin{figure}
\begin{center}
\vspace{-5mm}
\includegraphics[angle=0,width=0.35 \textwidth]{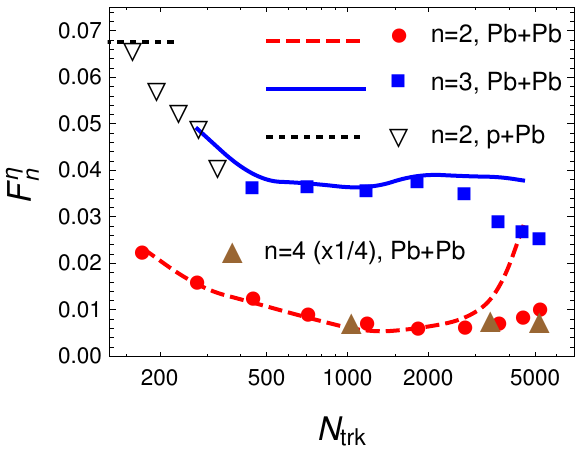}  
\end{center}
\vspace{-5mm}
\caption{Effective ``slope'' parameter $F^\eta_n$ of Eq.~(\ref{eq:F}) plotted as a function of the number of charged tracks. The
 data from the CMS collaboration (symbols)~\cite{Khachatryan:2015oea} are compared to calculations in the torque 
model with pseudorapidity fluctuating entropy density (lines) for $F^\eta_2$ and $F^\eta_3$. The data points for $F^\eta_4$
are scaled by $1/4$ (full triangles) to test the relation~(\ref{eq:F4}).
\label{fig:F}} 
\end{figure}

In Fig.~\ref{fig:F} we show the parameter $F^\eta_n$ of Eq.~(\ref{eq:F}) for different centralities. The calculation 
describes properly the measured  $r_2$ and  $r_3$ in  Pb-Pb collisions from semi-central to peripheral collisions.
In central and ultra-central collisions the model overestimates the factorization breaking, especially for $r_2$.
As stated above, this indicates that in central collisions other sources of correlations appear that  are not captured in our 
model. The calculation reproduces $r_2$ measured in p-Pb collisions, but not its centrality dependence.
The experimental data  for the fourth order harmonic flow  $F^\eta_4$, scaled by a factor $1/4$
are very close to the numbers for $F^\eta_2$. This is in agreement with relation (\ref{eq:F4}), and is consistent 
with the collective flow scenario. We stress that this relation is independent of the specific 
model of  initial conditions, and holds only under the assumption that collective flow with large values of $v_2$
 is generated.

\section{Conclusions} 
 
In this paper the factorization breaking for event-plane angles defined at different pseudorapidities has been investigated
with the help of the factorization ratio coefficients $r_n$~\cite{Khachatryan:2015oea}. The observed factorization breaking  
confirms qualitatively the existence of event-plane decorrelation, as suggested already in Ref.~\cite{Bozek:2010vz}.
We have shown that the strength of the factorization breaking is a sensitive measure of the fluctuations of entropy deposition 
in space-time rapidity. The CMS data for p-Pb collisions strongly suggest the existence of such
fluctuations, that are independent  for each wounded nucleon. 
We have studied a simple model of fluctuating entropy distribution 
in space-time rapidity, where the entropy production profile from wounded nucleons is approximately uniform, but  
the position of its end-point in pseudorapidity fluctuates.  Calculations within the proposed 
torque model with fluctuating entropy distribution
describe fairly well the measurements of $r_2$ and $r_3$ in Pb-Pb collisions, except for central collisions.
The fluctuations in the pseudorapidity profile of the initial fireball are absolutely essential in reproducing the data for 
p-Pb collisions.

We have shown on general grounds that the rank-4 slope coefficient $F^\eta_4$ is very close to $4 F^\eta_2$, which is 
confirmed by the data and which is one more signature of collectivity in the fireball evolution. 

We thus argue, based on our analysis, that the measurement of CMS collaboration of the factorization breaking for different 
pseudorapidities demonstrates the existence of fluctuations in the initial fireball density.
A successful description of the observed  collective flow requires 
the introduction of realistic fluctuating entropy distributions in the initial stage of  the hydrodynamic evolution.
The proposed mechanism is probably not unique, and it would be very interesting to have similar estimates
from the  color glass condensate approach~\cite{McLerran:1993ni}. Further, more accurate simulations
should involve full 3+1-dimensional hydrodynamic evolution and estimates of non flow-correlations.

\bigskip
                               
Research supported by the Polish Ministry of Science and Higher Education (MNiSW), by the
National Science Center grants DEC-2012/05/B/ST2/02528 and DEC-2012/06/A/ST2/00390.

\bigskip

\bibliography{../hydr}

\end{document}